\documentclass[aps,prl,twocolumn,superscriptaddress,showpacs]{revtex4}

\usepackage{graphicx}
\usepackage{subfig}
\usepackage{dcolumn}
\usepackage{bm}
\usepackage{times}

\def\ttbar{t\bar{t}}
\def\qqbar{q\bar{q}}
\def\ppbar{p\bar{p}}
\def\mMEt{\not\!\!{E_T}}
\def\MEt{\hbox{ $\mMEt$} }
\def\pt{$p_{T} \ $}
\def\et{$E_{T} \ $}
\def\avntrk{\mbox{$\langle N_{\mathrm{trk}}\rangle$}~}
\def\avntrkend{\mbox{$\langle N_{\mathrm{trk}}\rangle$}}
\def\avng{\mbox{$\langle N_g\rangle$}~}
\def\avngend{\mbox{$\langle N_g\rangle$}}

\begin{document}

\title{First Measurement of the Fraction of Top Quark Pair Production Through Gluon-Gluon Fusion}
\affiliation{Institute of Physics, Academia Sinica, Taipei, Taiwan 11529, Republic of China} 
\affiliation{Argonne National Laboratory, Argonne, Illinois 60439} 
\affiliation{Institut de Fisica d'Altes Energies, Universitat Autonoma de Barcelona, E-08193, Bellaterra (Barcelona), Spain} 
\affiliation{Baylor University, Waco, Texas  76798} 
\affiliation{Istituto Nazionale di Fisica Nucleare, University of Bologna, I-40127 Bologna, Italy} 
\affiliation{Brandeis University, Waltham, Massachusetts 02254} 
\affiliation{University of California, Davis, Davis, California  95616} 
\affiliation{University of California, Los Angeles, Los Angeles, California  90024} 
\affiliation{University of California, San Diego, La Jolla, California  92093} 
\affiliation{University of California, Santa Barbara, Santa Barbara, California 93106} 
\affiliation{Instituto de Fisica de Cantabria, CSIC-University of Cantabria, 39005 Santander, Spain} 
\affiliation{Carnegie Mellon University, Pittsburgh, PA  15213} 
\affiliation{Enrico Fermi Institute, University of Chicago, Chicago, Illinois 60637} 
\affiliation{Comenius University, 842 48 Bratislava, Slovakia; Institute of Experimental Physics, 040 01 Kosice, Slovakia} 
\affiliation{Joint Institute for Nuclear Research, RU-141980 Dubna, Russia} 
\affiliation{Duke University, Durham, North Carolina  27708} 
\affiliation{Fermi National Accelerator Laboratory, Batavia, Illinois 60510} 
\affiliation{University of Florida, Gainesville, Florida  32611} 
\affiliation{Laboratori Nazionali di Frascati, Istituto Nazionale di Fisica Nucleare, I-00044 Frascati, Italy} 
\affiliation{University of Geneva, CH-1211 Geneva 4, Switzerland} 
\affiliation{Glasgow University, Glasgow G12 8QQ, United Kingdom} 
\affiliation{Harvard University, Cambridge, Massachusetts 02138} 
\affiliation{Division of High Energy Physics, Department of Physics, University of Helsinki and Helsinki Institute of Physics, FIN-00014, Helsinki, Finland} 
\affiliation{University of Illinois, Urbana, Illinois 61801} 
\affiliation{The Johns Hopkins University, Baltimore, Maryland 21218} 
\affiliation{Institut f\"{u}r Experimentelle Kernphysik, Universit\"{a}t Karlsruhe, 76128 Karlsruhe, Germany} 
\affiliation{Center for High Energy Physics: Kyungpook National University, Daegu 702-701, Korea; Seoul National University, Seoul 151-742, Korea; Sungkyunkwan University, Suwon 440-746, Korea; Korea Institute of Science and Technology Information, Daejeon, 305-806, Korea; Chonnam National University, Gwangju, 500-757, Korea} 
\affiliation{Ernest Orlando Lawrence Berkeley National Laboratory, Berkeley, California 94720} 
\affiliation{University of Liverpool, Liverpool L69 7ZE, United Kingdom} 
\affiliation{University College London, London WC1E 6BT, United Kingdom} 
\affiliation{Centro de Investigaciones Energeticas Medioambientales y Tecnologicas, E-28040 Madrid, Spain} 
\affiliation{Massachusetts Institute of Technology, Cambridge, Massachusetts  02139} 
\affiliation{Institute of Particle Physics: McGill University, Montr\'{e}al, Canada H3A~2T8; and University of Toronto, Toronto, Canada M5S~1A7} 
\affiliation{University of Michigan, Ann Arbor, Michigan 48109} 
\affiliation{Michigan State University, East Lansing, Michigan  48824} 
\affiliation{University of New Mexico, Albuquerque, New Mexico 87131} 
\affiliation{Northwestern University, Evanston, Illinois  60208} 
\affiliation{The Ohio State University, Columbus, Ohio  43210} 
\affiliation{Okayama University, Okayama 700-8530, Japan} 
\affiliation{Osaka City University, Osaka 588, Japan} 
\affiliation{University of Oxford, Oxford OX1 3RH, United Kingdom} 
\affiliation{University of Padova, Istituto Nazionale di Fisica Nucleare, Sezione di Padova-Trento, I-35131 Padova, Italy} 
\affiliation{LPNHE, Universite Pierre et Marie Curie/IN2P3-CNRS, UMR7585, Paris, F-75252 France} 
\affiliation{University of Pennsylvania, Philadelphia, Pennsylvania 19104} 
\affiliation{Istituto Nazionale di Fisica Nucleare Pisa, Universities of Pisa, Siena and Scuola Normale Superiore, I-56127 Pisa, Italy} 
\affiliation{University of Pittsburgh, Pittsburgh, Pennsylvania 15260} 
\affiliation{Purdue University, West Lafayette, Indiana 47907} 
\affiliation{University of Rochester, Rochester, New York 14627} 
\affiliation{The Rockefeller University, New York, New York 10021} 
\affiliation{Istituto Nazionale di Fisica Nucleare, Sezione di Roma 1, University of Rome ``La Sapienza," I-00185 Roma, Italy} 
\affiliation{Rutgers University, Piscataway, New Jersey 08855} 
\affiliation{Texas A\&M University, College Station, Texas 77843} 
\affiliation{Istituto Nazionale di Fisica Nucleare, University of Trieste/\ Udine, Italy} 
\affiliation{University of Tsukuba, Tsukuba, Ibaraki 305, Japan} 
\affiliation{Tufts University, Medford, Massachusetts 02155} 
\affiliation{Waseda University, Tokyo 169, Japan} 
\affiliation{Wayne State University, Detroit, Michigan  48201} 
\affiliation{University of Wisconsin, Madison, Wisconsin 53706} 
\affiliation{Yale University, New Haven, Connecticut 06520} 
\author{T.~Aaltonen}
\affiliation{Division of High Energy Physics, Department of Physics, University of Helsinki and Helsinki Institute of Physics, FIN-00014, Helsinki, Finland}
\author{J.~Adelman}
\affiliation{Enrico Fermi Institute, University of Chicago, Chicago, Illinois 60637}
\author{T.~Akimoto}
\affiliation{University of Tsukuba, Tsukuba, Ibaraki 305, Japan}
\author{M.G.~Albrow}
\affiliation{Fermi National Accelerator Laboratory, Batavia, Illinois 60510}
\author{B.~\'{A}lvarez~Gonz\'{a}lez}
\affiliation{Instituto de Fisica de Cantabria, CSIC-University of Cantabria, 39005 Santander, Spain}
\author{S.~Amerio}
\affiliation{University of Padova, Istituto Nazionale di Fisica Nucleare, Sezione di Padova-Trento, I-35131 Padova, Italy}
\author{D.~Amidei}
\affiliation{University of Michigan, Ann Arbor, Michigan 48109}
\author{A.~Anastassov}
\affiliation{Rutgers University, Piscataway, New Jersey 08855}
\author{A.~Annovi}
\affiliation{Laboratori Nazionali di Frascati, Istituto Nazionale di Fisica Nucleare, I-00044 Frascati, Italy}
\author{J.~Antos}
\affiliation{Comenius University, 842 48 Bratislava, Slovakia; Institute of Experimental Physics, 040 01 Kosice, Slovakia}
% ...
\collaboration{CDF Collaboration\footnote{With visitors from $^a$University of Athens, 15784 Athens, Greece, 
$^b$Chinese Academy of Sciences, Beijing 100864, China, 
$^c$University of Bristol, Bristol BS8 1TL, United Kingdom, 
$^d$University Libre de Bruxelles, B-1050 Brussels, Belgium, 
$^e$University of California Irvine, Irvine, CA  92697, 
$^f$University of California Santa Cruz, Santa Cruz, CA  95064, 
$^g$Cornell University, Ithaca, NY  14853, 
$^h$University of Cyprus, Nicosia CY-1678, Cyprus, 
$^i$University College Dublin, Dublin 4, Ireland, 
$^j$University of Edinburgh, Edinburgh EH9 3JZ, United Kingdom, 
$^k$University of Heidelberg, D-69120 Heidelberg, Germany, 
$^l$Universidad Iberoamericana, Mexico D.F., Mexico, 
$^m$University of Manchester, Manchester M13 9PL, England, 
$^n$Nagasaki Institute of Applied Science, Nagasaki, Japan, 
$^o$University de Oviedo, E-33007 Oviedo, Spain, 
$^p$Queen Mary, University of London, London, E1 4NS, England, 
$^q$Texas Tech University, Lubbock, TX  79409, 
$^r$IFIC(CSIC-Universitat de Valencia), 46071 Valencia, Spain, 
}}
\noaffiliation

\begin{abstract}
We present the first measurement of $\sigma(gg \rightarrow t\bar{t})/\sigma(p\bar{p} \rightarrow t\bar{t})$. We use 0.96~fb$^{-1}$ of $\sqrt s =$1.96~TeV $\ppbar$ collision data recorded with the CDF II detector at Fermilab. We identify the candidate $\ttbar$ events with a high-energy charged lepton, a neutrino candidate, and four or more jets. Using charged particles with low transverse momentum in $\ttbar$ events, we find \mbox{$\sigma(gg \rightarrow t\bar{t})/\sigma(p\bar{p} \rightarrow t\bar{t})$}=0.07$\pm$0.14(stat)$\pm$0.07(syst), corresponding to a 95\% confidence level upper limit of 0.33, in agreement with the standard model NLO prediction of $0.15\pm0.05$.
\end{abstract}


\maketitle

Many studies have been dedicated to the understanding of the top quark, motivated in part by its large mass that may give it a unique role in the generation of mass for the quarks, leptons, and force carriers in the standard model (SM) of particle physics. In $\ppbar$ collisions at a center-of-momentum energy of $\sqrt{s}=$1.96~TeV, $(15 \pm 5)\%$ of $\ttbar$ pairs are expected to be produced through gluon-gluon fusion and the rest through quark-antiquark annihilation~\cite{ref: pdf sys, ref: xstn theo}, based on  next-to-leading-order (NLO) quantum chromodynamics (QCD) calculations. The inclusive $\ttbar$ production cross section has been measured by both CDF~\cite{ref: xtn PRD, ref: xtn meas CDF} and D0~\cite{ref: xtn meas D0} collaborations using various methods and decay modes of the $\ttbar$ pairs, and the results are in agreement with SM predictions. However, the details of the production process have never been investigated.

A measurement of the $gg \rightarrow \ttbar$ production cross section tests the QCD prediction. Also of interest is any indication of new top quark production and decay mechanisms. As the partonic cross section calculations are directly related to the momentum distributions of constituents of the colliding protons~\cite{ref: pdf sys}, such a measurement could assist in reducing the uncertainties in the gluon distributions within protons.

In this Letter we report the first measurement of the fractional cross section of $\ttbar$ production through gluon-gluon fusion, $\sigma(gg \rightarrow \ttbar)/\sigma(\ppbar \rightarrow \ttbar)$. To discriminate between the similar final state signatures of $gg \rightarrow \ttbar$ and $\qqbar \rightarrow \ttbar$, we take advantage of the higher probability for a gluon than for a quark to radiate a low-momentum gluon~\cite{GLAP}. Therefore, on average we expect a larger low-momentum charged particle multiplicity in $gg \rightarrow \ttbar$ compared to $\qqbar \rightarrow \ttbar$. Given the large theoretical uncertainties associated with gluon radiation, we do not rely on theoretical calculations for the modeling of the charged particle multiplicity. Instead, we use two different processes, $W$+n jet and two-jet (dijet) production, with well-understood production mechanisms, as calibration samples to relate the observed charged particle multiplicity to the fraction of processes involving more gluons.

We use a data sample of $\sqrt s =$1.96~TeV $\ppbar$ collisions with an integrated luminosity of 0.96$\pm$0.06~fb$^{-1}$ recorded by the CDF II detector at Fermilab between March 2002 and February 2006. The CDF II detector is described in detail in~\cite{CDF}; here, we briefly discuss the components essential for this analysis. The detector consists of a tracking system immersed in a solenoidal magnetic field of 1.4~T and electromagnetic and hadronic calorimeters surrounding the solenoid, followed by the muon system. Electrons, photons, and hadronic jets are identified using calorimeters and the tracking information. Muons are identified by the muon system together with tracking and calorimeter information. The data are collected using a three-level trigger system.

According to the SM top quarks almost always decay to a $W$ boson and a bottom quark, and so in $\ttbar$ events we expect to have two $W$ bosons and two $b$ quarks. We select $\ttbar$ candidate events where one of the $W$ bosons decays to two jets and the other decays to a lepton ($l$) and the corresponding neutrino.  In this analysis $l$ is either an electron or a muon. Our first calibration dataset is a set of $W$($\rightarrow l \nu$)+n jet (n = 0, 1, 2, 3) candidate events, for which the number of gluons involved in the production process increases with the number of jets~\cite{Wnjet pro}. The second is a set of events with two back-to-back, high-energy jets.  The average number of gluons involved in dijet production~\cite{dijet pro} falls with increasing transverse energy ($E_{T}$)~\cite{coord} of the highest $E_T$ jet (leading jet), as the relative rate of the $qq \rightarrow qq$, $qg \rightarrow qg$ and $gg \rightarrow gg$ subprocesses change. The number of gluons in each subprocess is 0, 2, and 4 respectively.

The $W$+ n jet data are collected with an inclusive lepton trigger that requires an electron with $E_T>$18~GeV or a muon with $p_T>$18~GeV/$c$. We select events with a reconstructed isolated electron (muon) candidate with $E_T > 20$~GeV ($p_T > 20$~GeV/$c$) and a missing \et (\MEt) $> 20$~GeV. We categorize the $W$+n jet samples by n, the number of jet candidates with $E_T>$15~GeV and pseudorapidity region $| \eta | <$ 2. Jets are defined using an iterative cone algorithm~\cite{ref:JetClu} with a cone of $\Delta R = \sqrt{(\Delta \phi)^2 + (\Delta \eta)^2} = $0.4 and are corrected for absolute energy response, $\eta$ dependence of calorimeter response, and multiple interactions. For the $\ttbar$ data sample, in addition to the above, we require four or more jets where at least one is identified as originating from a $b$ quark ($b$ tag). To define a $b$ tag, we identify within a jet a long-lived $B$-hadron candidate through the presence of a displaced secondary vertex~\cite{ref: xtn PRD}. In both $\ttbar$ and $W$+n jet samples, we remove any event with a second lepton candidate consistent with arising from a $Z$ boson decay or a $\ttbar$ event in which both $W$ bosons decay to leptons. We also veto the events in which the electron (muon) is consistent with coming from a conversion photon (cosmic ray)~\cite{ref: xtn PRD}. The dijet data are collected using two inclusive jet triggers that require a jet with \et of at least 50~GeV or at least 100~GeV (Jet50 and Jet100 datasets). We require a minimum leading jet \et of 75 and \mbox{130~GeV} for Jet50 and Jet100 datasets, respectively, to avoid any trigger bias. We remove events containing an electron (muon) candidate with $E_T>$20~GeV ($p_T>$20~GeV/$c$). We also require exactly two jets with \mbox{$|\eta| \leq$ 2} and a minimum \et of \mbox{20~GeV} and with the two jets back-to-back, having a $|\Delta \phi| \geq 2.53$~rad.

The background processes in our $\ttbar$ sample consist of $W$+ jets, electroweak processes ($WW$, $WZ$, $ZZ$), single top quark, and multi-jet QCD processes (non-$W$). For non-$W$ and $W$+ jets background, we can have a real $b$ tag (heavy flavor background, HF) or have a $b$ tag due to misidentification (light flavor background, LF). We estimate LF and HF in events with a real $W$ boson using various calibration datasets. For the small fraction of events from non-$W$ sources, we assume the non-$W$ background is equal parts HF and LF.  The results of the analysis are insensitive to this assumption. Single top quark processes are part of HF, while diboson backgrounds, ignoring the few $Z \rightarrow b\bar{b}$ events, are included in LF. We find 240 $\ttbar$ candidates with an estimated background contamination of (13$\pm$2)\%. The background estimates are found using the method explained in~\cite{ref: xtn PRD}.

The number of low-$p_T$ charged particles $N_{\mathrm{trk}}$ is affected by low-energy particles arising from jet fragmentation as well as multiple interactions within the same $\ppbar$ bunch crossing.  To include a track in our definition of $N_{\mathrm{trk}}$, we require it to have a \pt in the range 0.3-2.9~GeV/$c$ and $|\eta| \leq$ 1.1 and to originate from the vertex associated with the charged leptons and jets. We reject the track if it falls within $\Delta R =$0.6 and $\Delta R =$0.4 of jets with \et $\geq$15~GeV (high \et jets) and 6$\leq E_{T} <$15~GeV (low \et jets), respectively. Excluding these tracks results in different available tracking area for each event. We therefore correct the observed multiplicity to the total tracking coverage in $\eta$ and $\phi$ event-by-event. The resulting track multiplicity still has a modest dependence on the number of high \et jets in the event. We therefore make a further correction to $N_{\mathrm{trk}}$ by measuring this dependence in multi-jet QCD candidate events and using this as a per-jet correction($\sim$ 1 track per jet) to the multiplicity for all jets with $|\eta| \leq$ 1.1.

\begin{figure}
\begin{center}
\includegraphics[width=2.2in]{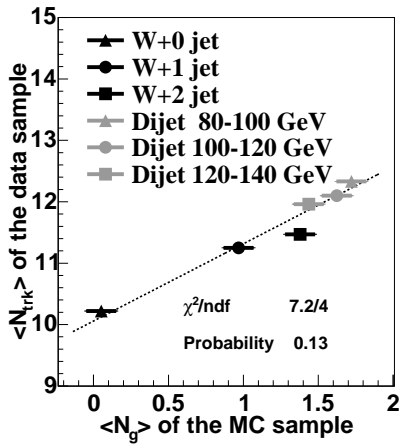}
\end{center}
\caption{The correlation between the average number of low-\pt charged particles (data) and the average number of gluons (MC). The dotted line is from a linear fit to the points.}
\label{fig: correlation}
\end{figure}

We show that there is a correlation between the average number of low-\pt charged particles \avntrk and the average number of gluons involved in the production process \avng in a given sample. We count the number of gluons that are part of the production process using Monte Carlo (MC) calculations for both dijet and the $W$+ n jet data samples. The $W$+ n jet MC sample is created using the \mbox{{\sc alpgen}~\cite{alpgen}} program followed by \mbox{{\sc pythia}~\cite{pythia}} to perform the jet fragmentation. The MC dijet events are created using the {\sc pythia} MC. We plot the observed \avntrk in data against the expected \avng from MC calculations for the calibration samples in Fig.~\ref{fig: correlation}. This demonstrates an approximately linear dependence between \avntrk and \avngend. We do not use this plot to obtain our result, but rather directly fit the observed $N_{\mathrm{trk}}$ distributions as described below.

The \avntrk and \avng correlation enables us to define $N_{\mathrm{trk}}$ distributions each representing a specific average number of gluons involved in the production process. We use this correlation and the observed $N_{\mathrm{trk}}$ distributions in the $W$+ 0 jet sample and the dijet sample with leading jet \et of 80-100~GeV to define a no-gluon and a gluon-rich $N_{\mathrm{trk}}$ distribution, respectively. To do so, an iterative procedure is adopted in order to remove the no-gluon (gluon-rich) contribution from the $N_{\mathrm{trk}}$ distribution of the 80-100 GeV dijet ($W$+ 0 jet) sample. There are no significant changes in the distributions after the first iteration. The $W$+ 0 jet sample is largely composed of the Drell-Yan $q\bar{q}^{\prime}$ process with a small QCD background of order 4\% and contribution from $W$ production in association with other partons where none of the final state jets are detected. The fraction of $W$+ 0 jet candidates with production processes involving gluons is estimated to be ($5\pm4$)\%. The no-gluon contribution of dijet candidates with leading jet \et of 80-100~GeV comes from $qq \rightarrow qq$ processes and is estimated to be ($27\pm3$)\%. 

To verify that the no-gluon or gluon-rich distribution can model the $N_{\mathrm{trk}}$ distribution of any process with comparable \avng regardless of the center-of-momentum energy, we check the $W$+ 1 jet data sample, and we see no dependence on jet \et in \avntrkend. 

The gluon-rich fraction associated with a given $N_{\mathrm{trk}}$ distribution can be found using a binned likelihood fit of the $N_{\mathrm{trk}}$ distribution of the form

\begin{equation}
N [ f_{g} \mathcal{F}_{g}(N_{\mathrm{trk}}) + (1 - f_{g}) \mathcal{F}_{q}(N_{\mathrm{trk}}) ],
\end{equation}

\noindent where $N$ is the normalization factor and one of the free parameters, $f_{g}$ is the fraction of gluon-rich components of the sample and the second free parameter, and $\mathcal{F}_{g}(N_{\mathrm{trk}})$ and $\mathcal{F}_{q}(N_{\mathrm{trk}})$ are the normalized gluon-rich and no-gluon parameterizations, respectively. We have verified this technique is free of bias using randomly generated samples from the data $N_{\mathrm{trk}}$ distributions.

\begin{figure}
\begin{center} 
\includegraphics[width=2.2in]{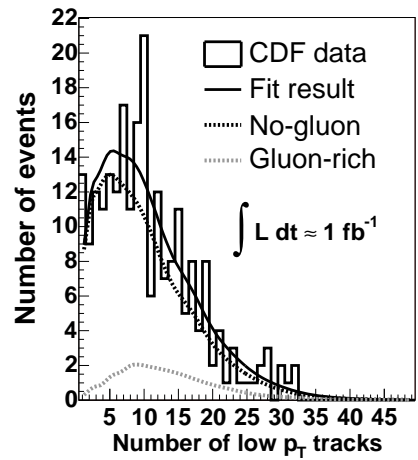}
\end{center}
\caption{The number of low-\pt charged particles for the $\ttbar$ candidates, the fit result, the gluon-rich, and no-gluon components.}
\label{fig: W4_tag}
\end{figure}

The $N_{\mathrm{trk}}$ distribution of the $\ttbar$ candidates, shown in Fig.~\ref{fig: W4_tag}, has a mean of 10.6$\pm$0.5. The fit, shown in the figure, models the data distribution very well, based on a goodness of fit test with 92\% probability. The measured gluon-rich fraction in $\ttbar$ candidates determined by fitting the $N_{\mathrm{trk}}$ distribution consists of two components, the $\ttbar$ gluon-rich fraction and the background gluon-rich fraction. Therefore, knowing the background fraction in our sample $f_{b}$ and the measured $f_{g}$ from the fit, we can write

\begin{equation}
f_{g} = f_{b}f_{g}^{\mathrm{bkg}} + (1 - f_{b})f_{g}^{\ttbar},
\end{equation}

\noindent where $f_{g}^{\mathrm{bkg}}$ and $f_{g}^{\ttbar}$ are the gluon-rich fraction of the background and $\ttbar$ signal, respectively. To estimate $f_{g}^{\mathrm{bkg}}$, we measure $f_{g}$ in the $W$+ 1, $W$+ 2, and $W$+ 3 jet data samples with no $b$ tag and with at least one $b$ tag using the fit to the N$_{trk}$ distribution for each of these samples. We then extrapolate the $f_{g}$ values from the $W$+ 1, 2, and 3 jet samples to $W$+ 4 or more jet bins for both $b$-tag and no $b$-tag samples. We consider the $b$-tag sample as representative of HF and the no $b$-tag sample as representative of LF. Using these extrapolations and the LF and HF fractions, we find $f_{g}^{\mathrm{bkg}} = 0.54 \pm 0.09$ and $f_{g}^{\ttbar} = 0.09 \pm 0.16$.

Given $f_{g}^{\ttbar}$, we measure $\sigma(gg \rightarrow \ttbar)/\sigma(\ppbar \rightarrow \ttbar)$ as

\begin{equation}
\left[1-\frac{\mathcal{A}_{gg}}{\mathcal{A}_{q\bar{q}}} +\left(\frac{\mathcal{A}_{gg}}{\mathcal{A}_{q\bar{q}}}\right)\left(\frac{1}{f_{g}^{\ttbar}}\right)\right]^{-1}=0.07\pm0.14{\rm (stat)},
\end{equation}

\noindent where $\mathcal{A}_{gg}$ and $\mathcal{A}_{q\bar{q}}$ are the acceptance for $gg \rightarrow \ttbar$ and $q\bar{q} \rightarrow \ttbar$, respectively. Using {\sc pythia}~\cite{pythia} MC calculations, we find $(14.1 \pm 0.5)$\% and $(11.5 \pm 0.4)$\% for $\mathcal{A}_{gg}$ and $\mathcal{A}_{q\bar{q}}$, respectively. The acceptance uncertainties include the systematic uncertainties.

The systematic uncertainties of this measurement, a total of 0.07, arise from uncertainties in the measurement of $N_{\mathrm{trk}}$ and the subsequent calculations. The uncertainties in $N_{\mathrm{trk}}$ are due to the per-jet correction(0.05), the estimated gluon content of the $W$+ 0 jet sample(0.04), and the choice of the low \et jet cut(0.02). In addition to these sources, there are uncertainties associated with the estimated $qq \rightarrow qq$ fraction of the 80-100~GeV dijet sample, the background fraction, the modeling of the background gluon-rich fraction, the non-$W$ background fraction, and the acceptances; these are all negligible. To estimate the effects of all the above uncertainties, we changed the central values and measured the change in the relevant variables. 

The result corresponds to an upper limit of 0.33 at 95\% confidence level. We use a classical statistical technique to set the limit by simulating the possible outcomes for a given true value taking into account the systematic effects.

In conclusion, we have presented the first measurement of $\sigma(gg \rightarrow \ttbar)/\sigma(\ppbar \rightarrow \ttbar)$ and found 0.07$\pm$0.14(stat)$\pm$0.07(syst), corresponding to an upper limit of 0.33 at 95\% confidence level, in 0.96~fb$^{-1}$ of data collected at CDF. This is in agreement with the SM prediction of $0.15 \pm 0.05$, and does not suggest that non-standard model processes~\cite{ref: technicolor} contribute to top quark pair production at the Tevatron. 

We thank the Fermilab staff and the technical staffs of the participating institutions for their vital contributions. This work was supported by the U.S. Department of Energy and National Science Foundation; the Italian Istituto Nazionale di Fisica Nucleare; the Ministry of Education, Culture, Sports, Science and Technology of Japan; the Natural Sciences and Engineering Research Council of Canada; the National Science Council of the Republic of China; the Swiss National Science Foundation; the A.P. Sloan Foundation; the Bundesministerium f\"ur Bildung und Forschung, Germany; the Korean Science and Engineering Foundation and the Korean Research Foundation; the Science and Technology Facilities Council and the Royal Society, UK; the Institut National de Physique Nucleaire et Physique des Particules/CNRS; the Russian Foundation for Basic Research; the Comisi\'on Interministerial de Ciencia y Tecnolog\'{\i}a, Spain; the European Community's Human Potential Programme; the Slovak R\&D Agency; and the Academy of Finland.


\begin{thebibliography}{99}
\bibitem{coord}
We use a coordinate system with an origin at the center of the detector, $\phi$ and $\theta$ as the azimuthal and polar angles, respectively, and a $z$ axis along the proton beam direction. The transverse energy is defined as \et$= E \sin{\theta}$ and particle momentum transverse to the beam is \pt$= p \sin{\theta}$. The missing \et (\MEt) is defined by $\not\!\! \vec{E}_T = - \sum_{i} E_T^i \hat{n}_i$, where $i$ is the calorimeter tower number with $|\eta| < 3.6$, and $\hat{n}_i$ is a unit vector perpendicular to the beam axis pointing at the $i^{th}$ calorimeter tower. We also define $\not\!\!E_T=|\not\!\!\vec{E}_T|$.
\end{thebibliography}
\end{document}